\documentclass[preprint,aps]{revtex4}
\usepackage{color}
\begin{document}
\preprint{TUHEP-TH-07158}
 \draft
\title{Electroweak Chiral Lagrangian for Left-right Symmetric Models}

\author{Ying Zhang$^1$, Shun-Zhi Wang$^1$, Feng-Jun Ge$^1$, Qing Wang$^{1,2}$}

\address{$^1$Department of Physics,Tsinghua University,Beijing 100084,P.R.China\footnote{mailing address} \\
    $^2$Center for High Energy Physics, Tsinghua University, Beijing 100084, P.R.china}

\date{April 17, 2007}

\begin{abstract}
 The complete list of electroweak chiral Lagrangian up to order of $p^4$ for left-right symmetric models
 with a neutral light higgs is provided. The connection of these operators to left and right
gauge boson mixings and masses is made and their contribution to
conventional generalized electroweak chiral Lagrangian with a
neutral light higgs included in is estimated.

\bigskip
PACS number(s): 12.60-i, 11.30.Qc, 11.30.Rd, 13.10.+q
\end{abstract}
\maketitle


\vspace{1cm}
Although Standard Model (SM) is proved to be very successful,  higgs
particle is missing in the real world which implies that electroweak
symmetry breaking  mechanism (EWSBM) is still not known. SM provides
us a version of EWSBM through higgs boson which suffers from
triviality and unnaturalness problems. Beyond SM, various new
physics models are invented which exhibite many alternative EWSBMs.
A class of  them are  models with muti-higgs bosons and  the common
feature of these models is the existence of a light neutral higgs
boson, this light higgs is responsible for unitarizing the
longitudinal electroweak gauge boson scattering amplitudes. The most
general description of this class of models is a non-linear realized
chiral Lagrangian with a light higgs included in. This chiral
Lagrangian is already written down in Ref.\cite{WangLiMing} and it
is formally equivalent to linearized version of chiral Lagrangian
with higgs and three goldstone bosons which form a complex doublet
representation of electroweak symmetry. It is the generalization of
conventional well-known nonlinear version of electroweak chiral
Lagrangian (EWCL) \cite{EWCL0,EWCL} by adding in theory a neutral
higgs and was called extended electroweak chiral Lagrangian (EEWCL)
in Ref.\cite{WangLiMing} which offers the most general description
of electro-weak interaction with a neutral higgs at low energy
region. The symmetry realization pattern for EEWCL is $SU(2)_{\rm
L}\otimes U(1)_{\rm Y}\rightarrow U(1)_{\rm em}$ and bosonic fields
of the theory involving electroweak interaction are higgs field $h$,
photon field $A$, weak gauge fields $W^{\pm}$, $Z^0$ and three
goldstone bosons which will be eaten out to become longitudinal
components of $W^{\pm},Z^0$ and then give them masses due to Higgs
mechanism. Within EEWCL, new physics models with light higgs at low
energy region can be parameterized by a set of coefficients, it
universally describes all possible electro-weak interactions among
existing particles and undiscovered neutral light higgs, it offers a
model independent platform for us to investigate various kinds
EWSBMs.

Except unknown EWSBM, another puzzle of electroweak interaction is
due to its chirality, the left-right non-symmetric interaction
 explicitly violate parity symmetry. The beauty of the symmetry and
 past successes of electromagnetic unification and electroweak
 unification driven some of us believe that the underlying elementary interaction
should be symmetric under parity transformation and then at certain
higher energy scale, parity symmetry should be recovered. This
inspires the idea of spontaneous parity violation (SPV) and people
then is interested in generalizing conventional electroweak
interaction model to left-right symmetric models of electroweak
interactions \cite{LRSMs}. If the interactions responsible for SPV
become strong at TeV energy region, we are possible to find
experiment evidences in next generation high energy colliders and
realize joint investigation both on EWSBM and SPV at TeV energy
region. Admit left-right symmetry requires doubling those particles
originally participate left hand electroweak interactions. So at
least the gauge fields involving electroweak interaction are doubled
by adding to original left handed $W^{\pm}_L$, $Z^0_L$ and three
corresponding goldstone bosons with their right handed ones. In this
paper we discuss a situation that beyond the particles already
discovered in past high energy experiments, the lightest new
particles are a light neutral higgs and right hand $W^{\pm}_R$,
$Z^0_R$ and their goldstone bosons. All other new particles are
heavier than right hand $W^{\pm}_R$, $Z^0_R$. Then below the
threshold of these more heavier particles, if we want to setup a
model independent description for all possible left-right symmetric
interactions, we are leading to the demand of building a chiral
Lagrangian for left-right symmetric models. It is the purpose of
this work to construct such a Lagrangian. This Lagrangian will be
the most general and economic low energy description of the various
left-right-symmetric models for the case that only one neutral higgs
is lighter than $W^{\pm}_R$, $Z^0_R$, since all different EWSBMs and
SPV mechanisms reside in the parameters of the Lagrangian which
enable us to evade the details of spontaneous symmetry breaking
mechanism. This Lagrangian will provide us larger parameter space
than any of detail left-right symmetric models, it will be more
flexible than detail models when we use it to compare with exiting
experiment data and then to perform more complete test for the
possibility of realization in nature that all new particles except
light higgs are heavier than $W^{\pm}_R$, $Z^0_R$. Even the final
phenomenological investigations find inconsistencies with experiment
data, which imply there does some extra particles beyond light higgs
lighter than $W^{\pm}_R$, $Z^0_R$, the formalism developed in this
work is still useful, the only change is that we need to add these
extra new particles into the Lagrangian. Consider there are two much
uncertainties for these extra particles, as the first step of the
model independent investigation, we consider the minimal particle
content required by left-right symmetry and limit the new particles
in this work only in the content of light neutral higgs, left and
right handed $W^{\pm}$, $Z^0$ and their goldstone bosons. The
possible generalizations will be discussed elsewhere. The symmetry
realization pattern now is generalized from original $SU(2)_{\rm
L}\otimes U(1)_{\rm Y}\rightarrow U(1)_{\rm em}$ to $SU(2)_{\rm
L}\otimes SU(2)_{\rm R}\otimes U(1)_{\rm B-L} \rightarrow SU(2)_{\rm
L}\otimes U(1)_{\rm Y}\rightarrow U(1)_{\rm em}$. The nonlinear
realization of this symmetry allows us to build up our theory with
only one neutral higgs boson. For simplicity, we only discuss
bosonic part of chiral Lagrangian in this work.

 Let $B_{\mu}$, $W_{L,\mu}^a$,
$W_{R,\mu}^a$ be electroweak gauge fields ($a=1,2,3$) and two by two
unitary unimodular matrices $U_L$ and $U_R$ be corresponding
goldstone boson fields, $h$ be neutral higgs field . Under
$SU(2)_L\otimes SU(2)_R\otimes U(1)_{B-L}$ transformations, higgs
field is invariant and other fields transform as,
\begin{eqnarray}
&&B_{\mu}(x)\;\rightarrow\;B_{\mu}(x)
-\frac{1}{g}[\partial_{\mu}\theta^0(x)]\;,\hspace{2cm}U_i(x)\;\rightarrow\;
e^{\frac{i}{2}\tau^a\theta^a_i(x)}U_i(x)
e^{-\frac{i}{2}\tau^3\theta^0(x)}\hspace{1cm}i=L,R\;,\nonumber\\
&&\frac{\tau^a}{2}W_{i,\mu}^a(x)\;\rightarrow\;e^{i\frac{\tau^b}{2}\theta^b_i(x)}\frac{\tau^a}{2}W_{i,\mu}^a(x)
e^{-i\frac{\tau^c}{2}\theta^c_i(x)}-\frac{i}{g_i}e^{i\frac{\tau^b}{2}\theta^b_i(x)}
\partial_{\mu}e^{-i\frac{\tau^c}{2}\theta^c_i(x)}\;.\label{su2LRtran}
\end{eqnarray}
Covariant derivative of goldstone fields are
\begin{eqnarray}
D_{\mu}U_i&=&\partial_{\mu}U_i+ig_i\frac{\tau^a}{2}W^a_{i,\mu}U_i-igU_i\frac{\tau_3}{2}B_{\mu}\hspace{2cm}i=L,R
\end{eqnarray}
Since naive generalization of conventional EWCL building blocks
$T_i=U_i\tau_3U^\dagger_i$ and $V_{i,\mu}=(D_{\mu}U_i)U_i^\dagger$
are not convenient for present discussion, we introduce alternative
building blocks
\begin{eqnarray}
&&B_{\mu\nu}=\partial_{\mu}B_{\nu}-\partial_{\nu}B_{\mu}\;,\hspace{2cm}X_i^\mu\equiv
U_i^\dag(D^\mu U_i)\;,\hspace{2cm}\overline{W}_{i,\mu\nu}\equiv
U_i^\dag
g_iW_{i,\mu\nu}U_i\;,\nonumber\\
&&W_{i,\mu\nu}=W_{i,\mu\nu}^a\frac{\tau^a}{2}=\partial_{\mu}W_{i,\nu}^a\frac{\tau^a}{2}-\partial_{\nu}W_{i,\mu}^a\frac{\tau^a}{2}
+ig_i[W_{i,\mu}^a\frac{\tau^a}{2},W_{i,\nu}^b\frac{\tau^b}{2}]\hspace{1cm}i=L,R\;.
\end{eqnarray}
Since the higgs field $h$ should develop vacuum expectation value,
we count it as order of $p^0$ in the power counting of the low
energy expansion. This imply the lowest order of chiral Lagrangian
is just the higgs potential
\begin{eqnarray}
\mathcal{L}_0&=&-V(h)
\end{eqnarray}
the next to leading order of the chiral Lagrangian then is $p^2$ of
Lagrangian,
\begin{eqnarray}
\mathcal{L}_2&=&\frac{1}{2}(\partial_{\mu}h)^2-\frac{1}{4}f_L^2{\rm
tr}(X_{L,\mu}X_L^{\mu})-\frac{1}{4}f_R^2{\rm
tr}(X_{R,\mu}X_R^{\mu})+\frac{1}{2}\kappa f_Lf_R{\rm tr}(X_L^{\mu}
X_R^{\mu})\\
&&+\frac{1}{4}\beta_{L,1}f_L^2[{\rm
tr}(\tau^3X_{L,\mu})]^2+\frac{1}{4}\beta_{R,1}f_L^2[{\rm
tr}(\tau^3X_{R,\mu})]^2+\frac{1}{2}\tilde{\beta}_1f_Lf_R[{\rm
tr}(\tau^3X_{L,\mu})][{\rm tr}(\tau^3X_R^{\mu})]\nonumber
\end{eqnarray}
in which $f_L,f_R,\beta_{L,1}, \beta_{R,1}, \tilde{\beta}_1$ are all
depend on higgs field $h$, but not depend on it derivative
$\partial_\mu h$ or $\partial^2h$, this is because these derivative
terms increase the powers of the low energy expansion. If we take
$h$ inside of $f_L$ and $f_R$ be its vacuum expectation value, the
result $f_L$ and $f_R$ take the scales of spontaneous breaking for
electroweak symmetry and parity respectively. $\beta_{L,1}$,
$\beta_{R,1}$ and $\tilde{\beta}_1$ are corresponding coefficients
responsible for $p^2$ order left,right and crossing custodial
symmetry violating interactions. Note the equations of motion will
lead $\partial_{\mu}\mathrm{tr}(\tau^3X^\mu_i)=0$ and then in
$\mathcal{L}_2$, there should be no terms like
$C_i(h)(\partial_{\mu}h)\mathrm{tr}(\tau^3X^\mu_i)$ $i=R,L$.

$p^4$ order Lagrangian can be divided into four parts
\begin{eqnarray}
\mathcal{L}_4=\mathcal{L}_K+\mathcal{L}_L+\mathcal{L}_{HL}+\mathcal{L}_R+\mathcal{L}_{HR}+\mathcal{L}_C
\end{eqnarray}
with kinetic part of $p^4$ order Lagrangian $\mathcal{L}_K$
\begin{eqnarray}
\mathcal{L}_K=-\frac{1}{4}W_{L,\mu\nu}^aW_L^{\mu\nu,a}-\frac{1}{4}W_{R,\mu\nu}^aW_R^{\mu\nu,a}-\frac{1}{4}B_{\mu\nu}B^{\mu\nu}
\label{LK}
\end{eqnarray}
and left(right) part of $p^4$ order Lagrangian without differential
of higgs $\mathcal{L}_i,~i=L,R$
\begin{eqnarray}
\mathcal{L}_i&=&\frac{1}{2}\alpha_{i,1}gB_{\mu\nu}{\rm
tr}(\tau^3\overline{W}_i^{\mu\nu}) +i\alpha_{i,2}gB_{\mu\nu}{\rm
tr}(\tau^3X^{\mu}_iX^{\nu}_i)
 +2i\alpha_{i,3}{\rm tr}(\overline{W}_{i,\mu\nu}X^{\mu}_iX^{\nu}_i)
 +\alpha_{i,4}[{\rm tr}(X_{i,\mu}X_{i,\nu})]^2
\nonumber\\
 && +\alpha_{i,5}[{\rm tr}(X_{i,\mu}^2)]^2
 +\alpha_{i,6}{\rm tr}(X_{i,\mu}X_{i,\nu}){\rm tr}(\tau^3X^{\mu}_i){\rm tr}(\tau^3X^{\nu}_i)
 +\alpha_{i,7}{\rm tr}(X_{i,\mu}^2)[{\rm tr}(\tau^3X_{i,\nu})]^2
 \nonumber\\
 &&+\frac{1}{4}\alpha_{i,8}[{\rm tr}(\tau^3\overline{W}_{i,\mu\nu})]^2
 +i\alpha_{i,9}{\rm tr}(\tau^3\overline{W}_{i,\mu\nu}){\rm tr}(\tau^3X_i^{\mu}X_i^{\nu})
 +\frac{1}{2}\alpha_{i,10}[{\rm tr}(\tau^3X_{i,\mu}){\rm tr}(\tau^3X_{i,\nu})]^2
 \nonumber\\
 &&+
 \alpha_{i,11}\epsilon^{\mu\nu\rho\lambda}{\rm tr}(\tau^3X_{i,\mu}){\rm tr}(X_{i,\nu}\overline{W}_{i,\rho\lambda})
 +2\alpha_{i,12}{\rm tr}(\tau^3X_{i,\mu}){\rm tr}(X_{i,\nu}\overline{W}_i^{\mu\nu})
 \nonumber\\
&&+\frac{1}{4}\alpha_{i,13}g\epsilon^{\mu\nu\rho\sigma}\!B_{\mu\nu}{\rm
tr}(\tau^3\overline{W}_{i,\rho\sigma}\!)
+\frac{1}{8}\alpha_{i,14}\epsilon^{\mu\nu\rho\sigma}{\rm
tr}(\tau^3\overline{W}_{i,\mu\nu}){\rm
tr}(\tau^3\overline{W}_{i,\rho\sigma})\;.
\end{eqnarray}
$\mathcal{L}_{Hi},~i=L,R$ are left(right) part of $p^4$ order
Lagrangian with differential of higgs
\begin{eqnarray}
\mathcal{L}_{Hi}&=&(\partial_{\mu}h)\{\bar{\alpha}_{Hi,1}\mathrm{tr}(\tau^3X_i^\mu)\mathrm{tr}(X_{i,\nu}^2)
+\bar{\alpha}_{Hi,2}\mathrm{tr}(\tau^3X_i^\nu)\mathrm{tr}(X_i^{\mu}X_{i,\nu})+
\bar{\alpha}_{Hi,3}\mathrm{tr}(\tau^3X_i^\nu)\mathrm{tr}(\tau^3X_i^{\mu}X_{i,\nu})\nonumber\\
&&+\bar{\alpha}_{Hi,4}\mathrm{tr}(\tau^3X_i^{\mu})[\mathrm{tr}(\tau^3X_{i,\nu})]^2
+i\bar{\alpha}_{Hi,5}\mathrm{tr}(\tau^3X_{i,\nu})\mathrm{tr}(\tau^3\overline{W}_i^{\mu\nu})
+
ig\bar{\alpha}_{Hi,6}B^{\mu\nu}\mathrm{tr}(\tau^3X_{i,\nu})\nonumber\\
&&+i\bar{\alpha}_{Hi,7}\mathrm{tr}(\tau^3\overline{W}_i^{\mu\nu}X_{i,\nu})
+i\bar{\alpha}_{Hi,8}\mathrm{tr}(\overline{W}_i^{\mu\nu}X_{i,\nu})\}
+(\partial_{\mu}h)(\partial_{\nu}h)[\bar{\alpha}_{Hi,9}\mathrm{tr}(\tau^3X_i^\mu)\mathrm{tr}(\tau^3X_i^\nu)\nonumber\\
&&+ \bar{\alpha}_{Hi,10}\mathrm{tr}(X_i^{\mu}X_i^\nu)]
+(\partial_{\mu}h)^2\{\bar{\alpha}_{Hi,11}[\mathrm{tr}(\tau^3X_{i,\nu})]^2
+ \bar{\alpha}_{Hi,12}\mathrm{tr}(X_{i,\nu}^2)\}\nonumber\\
&&+\bar{\alpha}_{Hi,13}(\partial_{\mu}h)^2(\partial_{\nu}h)\mathrm{tr}(\tau^3X_i^\nu)+\bar{\alpha}_{Hi,14}(\partial_{\mu}h)^4
\end{eqnarray}
The most complex interaction is the crossing part of $p^4$ order
Lagrangian
\begin{eqnarray}
\mathcal{L}_C&=&i\tilde{\alpha}_{2}gB_{\mu\nu}{\rm
tr}(\tau^3X^{\mu}_LX^{\nu}_R)
 +2i\tilde{\alpha}_{3,1}{\rm tr}(\overline{W}_{L,\mu\nu}X^{\mu}_RX^{\nu}_R)
+2i\tilde{\alpha}_{3,2}{\rm
tr}(\overline{W}_{R,\mu\nu}X^{\mu}_LX^{\nu}_L)
 \nonumber\\
&& +2i\tilde{\alpha}_{3,3}{\rm
 tr}(\overline{W}_{L,\mu\nu}X^{\mu}_LX^{\nu}_R)+2i\tilde{\alpha}_{3,4}{\rm
tr}(\overline{W}_{R,\mu\nu}X^{\mu}_RX^{\nu}_L)
 +\tilde{\alpha}_{4,1}{\rm tr}(X_{L,\mu}X_{L,\nu}){\rm tr}(X_R^{\mu}X_R^{\nu})
\nonumber\\
&&+\tilde{\alpha}_{4,2}[{\rm
tr}(X_{L,\mu}X_{R,\nu})]^2+\tilde{\alpha}_{4,3}{\rm
tr}(X_{L,\mu}X_{R,\nu}){\rm
tr}(X_R^{\mu}X_L^{\nu})+\tilde{\alpha}_{4,4}{\rm
tr}(X_{L,\mu}X_{R,\nu}){\rm
tr}(X_R^{\mu}X_R^{\nu})\nonumber\\
 &&+\tilde{\alpha}_{4,5}{\rm
tr}(X_{R,\mu}X_{L,\nu}){\rm tr}(X_L^{\mu}X_L^{\nu})
+\tilde{\alpha}_{5,1}{\rm tr}(X_{L,\mu}^2){\rm tr}(X_{R,\nu}^2)
+\tilde{\alpha}_{5,2}[{\rm
tr}(X_{L,\mu}X_R^{\mu})]^2 \nonumber\\
&&+\tilde{\alpha}_{5,3}{\rm tr}(X_{L,\mu}X_R^{\mu}){\rm
tr}(X_{R,\nu}^2) +\tilde{\alpha}_{5,4}{\rm
tr}(X_{R,\mu}X_L^{\mu}){\rm
tr}(X_{L,\nu}^2)\nonumber\\
&&+\tilde{\alpha}_{6,1}{\rm tr}(X_{L,\mu}X_{L,\nu}){\rm
tr}(\tau^3X^{\mu}_R){\rm
tr}(\tau^3X^{\nu}_R)+\tilde{\alpha}_{6,2}{\rm
tr}(X_{R,\mu}X_{R,\nu}){\rm tr}(\tau^3X^{\mu}_L){\rm
tr}(\tau^3X^{\nu}_L)
\nonumber\\
&&+\tilde{\alpha}_{6,3}{\rm tr}(X_{L,\mu}X_{R,\nu}){\rm
tr}(\tau^3X^{\mu}_L){\rm
tr}(\tau^3X^{\nu}_R)+\tilde{\alpha}_{6,4}{\rm
tr}(X_{L,\mu}X_{R,\nu}){\rm tr}(\tau^3X^{\mu}_R){\rm
tr}(\tau^3X^{\nu}_L)\nonumber\\
&&+\tilde{\alpha}_{6,5}{\rm tr}(X_{L,\mu}X_{R,\nu}){\rm
tr}(\tau^3X^{\mu}_R){\rm
tr}(\tau^3X^{\nu}_R)+\tilde{\alpha}_{6,6}{\rm
tr}(X_{R,\mu}X_{L,\nu}){\rm tr}(\tau^3X^{\mu}_L){\rm
tr}(\tau^3X^{\nu}_L)\nonumber\\
&&+\tilde{\alpha}_{6,7}{\rm tr}(X_{L,\mu}X_{L,\nu}){\rm
tr}(\tau^3X^{\mu}_L){\rm
tr}(\tau^3X^{\nu}_R)+\tilde{\alpha}_{6,8}{\rm
tr}(X_{R,\mu}X_{R,\nu}){\rm tr}(\tau^3X^{\mu}_R){\rm
tr}(\tau^3X^{\nu}_L)
 \nonumber\\
&&+\tilde{\alpha}_{7,1}{\rm tr}(X_{L,\mu}^2)[{\rm
tr}(\tau^3X_{R,\nu})]^2+\tilde{\alpha}_{7,2}{\rm
tr}(X_{R,\mu}^2)[{\rm tr}(\tau^3X_{L,\nu})]^2\nonumber\\
&&+\tilde{\alpha}_{7,3}{\rm tr}(X_{L,\mu}X_R^{\mu}){\rm
tr}(\tau^3X_{L,\nu}){\rm
tr}(\tau^3X_R^{\nu})+\tilde{\alpha}_{7,4}{\rm
tr}(X_{L,\mu}X_R^{\mu})[{\rm tr}(\tau^3X_{R,\nu})]^2\nonumber\\
&& +\tilde{\alpha}_{7,5}{\rm tr}(X_{R,\mu}X_L^{\mu})[{\rm
tr}(\tau^3X_{L,\nu})]^2+\tilde{\alpha}_{7,6}{\rm
tr}(X_{L,\mu}^2){\rm tr}(\tau^3X_{L,\nu}){\rm
tr}(\tau^3X_R^{\nu}) \nonumber\\
 &&+\tilde{\alpha}_{7,7}{\rm
tr}(X_{R,\mu}^2){\rm tr}(\tau^3X_{R,\nu}){\rm
tr}(\tau^3X_L^{\nu})+\frac{1}{4}\tilde{\alpha}_8{\rm
tr}(\tau^3\overline{W}_{L,\mu\nu}){\rm
tr}(\tau^3\overline{W}_R^{\mu\nu})\nonumber\\ &&
+i\tilde{\alpha}_{9,1}{\rm tr}(\tau^3\overline{W}_{L,\mu\nu}){\rm
tr}(\tau^3X_R^{\mu}X_R^{\nu}) +i\tilde{\alpha}_{9,2}{\rm
tr}(\tau^3\overline{W}_{R,\mu\nu}){\rm
tr}(\tau^3X_L^{\mu}X_L^{\nu})\nonumber\\
&& +i\tilde{\alpha}_{9,3}{\rm tr}(\tau^3\overline{W}_{L,\mu\nu}){\rm
tr}(\tau^3X_L^{\mu}X_R^{\nu}) +i\tilde{\alpha}_{9,4}{\rm
tr}(\tau^3\overline{W}_{R,\mu\nu}){\rm tr}(\tau^3X_R^{\mu}X_L^{\nu})
 \nonumber\\
&& +\frac{1}{2}\tilde{\alpha}_{10,1}[{\rm tr}(\tau^3X_{L,\mu}){\rm
tr}(\tau^3X_{R,\nu})]^2+\frac{1}{2}[\tilde{\alpha}_{10,2}[{\rm
tr}(\tau^3X_{L,\mu}){\rm tr}(\tau^3X_R^{\mu})]^2
\nonumber\\
&&+\frac{1}{2}\tilde{\alpha}_{10,3}{\rm tr}(\tau^3X_{L,\mu}){\rm
tr}(\tau^3X_R^{\mu})[{\rm
tr}(\tau^3X_{R,\nu})]^2+\frac{1}{2}\tilde{\alpha}_{10,4}{\rm
tr}(\tau^3X_{R,\mu}){\rm
tr}(\tau^3X_L^{\mu})[{\rm tr}(\tau^3X_{L,\nu})]^2\nonumber\\
&&+ \tilde{\alpha}_{11,1}\epsilon^{\mu\nu\rho\lambda}{\rm
tr}(\tau^3X_{L,\mu}){\rm tr}(X_{R,\nu}\overline{W}_{R,\rho\lambda})
+ \tilde{\alpha}_{11,2}\epsilon^{\mu\nu\rho\lambda}{\rm
tr}(\tau^3X_{R,\mu}){\rm tr}(X_{L,\nu}\overline{W}_{L,\rho\lambda})\nonumber\\
 &&+ \tilde{\alpha}_{11,3}\epsilon^{\mu\nu\rho\lambda}{\rm
tr}(\tau^3X_{L,\mu}){\rm tr}(X_{L,\nu}\overline{W}_{R,\rho\lambda})+
\tilde{\alpha}_{11,4}\epsilon^{\mu\nu\rho\lambda}{\rm
tr}(\tau^3X_{R,\mu}){\rm
tr}(X_{R,\nu}\overline{W}_{L,\rho\lambda})\nonumber\\ &&
+\tilde{\alpha}_{11,5}\epsilon^{\mu\nu\rho\lambda}{\rm
tr}(\tau^3X_{L,\mu}){\rm
tr}(X_{R,\nu}\overline{W}_{L,\rho\lambda})+\tilde{\alpha}_{11,6}\epsilon^{\mu\nu\rho\lambda}{\rm
tr}(\tau^3X_{R,\mu}){\rm tr}(X_{L,\nu}\overline{W}_{R,\rho\lambda})\nonumber\\
&& +2\tilde{\alpha}_{12,1}{\rm tr}(\tau^3X_{L,\mu}){\rm
 tr}(X_{R,\nu}\overline{W}_R^{\mu\nu})+2\tilde{\alpha}_{12,2}{\rm tr}(\tau^3X_{R,\mu}){\rm
tr}(X_{L,\nu}\overline{W}_L^{\mu\nu})\nonumber\\
&& +2\tilde{\alpha}_{12,3}{\rm tr}(\tau^3X_{L,\mu}){\rm
tr}(X_{L,\nu}\overline{W}_R^{\mu\nu})+2\tilde{\alpha}_{12,4}{\rm
tr}(\tau^3X_{R,\mu}){\rm
tr}(X_{R,\nu}\overline{W}_L^{\mu\nu}) \nonumber\\
&&+2\tilde{\alpha}_{12,5}{\rm tr}(\tau^3X_{L,\mu}){\rm
tr}(X_{R,\nu}\overline{W}_L^{\mu\nu})+2\tilde{\alpha}_{12,6}{\rm
tr}(\tau^3X_{R,\mu}){\rm
tr}(X_{L,\nu}\overline{W}_R^{\mu\nu})\nonumber\\
&& +\frac{1}{8}\tilde{\alpha}_{14}\epsilon^{\mu\nu\rho\sigma}{\rm
tr}(\tau^3\overline{W}_{L,\mu\nu}){\rm
tr}(\tau^3\overline{W}_{R,\rho\sigma})
+(\partial_{\mu}h)\{\tilde{\alpha}_{H,1,1}\mathrm{tr}(\tau^3X_R^\mu)\mathrm{tr}(X_{L,\nu}^2)\nonumber\\
&&+\tilde{\alpha}_{H,1,2}\mathrm{tr}(\tau^3X_L^\mu)\mathrm{tr}(X_{R,\nu}^2)
+\tilde{\alpha}_{H,2,1}\mathrm{tr}(\tau^3X_R^\nu)\mathrm{tr}(X_L^{\mu}X_{L,\nu})\nonumber\\
&&+\tilde{\alpha}_{H,2,2}\mathrm{tr}(\tau^3X_L^\nu)\mathrm{tr}(X_R^{\mu}X_{R,\nu})
+\tilde{\alpha}_{H,3,1}\mathrm{tr}(\tau^3X_R^\nu)\mathrm{tr}(\tau^3X_L^{\mu}X_{L,\nu})\nonumber\\
&&+\tilde{\alpha}_{H,3,2}\mathrm{tr}(\tau^3X_L^\nu)\mathrm{tr}(\tau^3X_R^{\mu}X_{R,\nu})
+\tilde{\alpha}_{H,4,1}\mathrm{tr}(\tau^3X_R^{\mu})[\mathrm{tr}(\tau^3X_{L,\nu})]^2\nonumber\\
&&+\tilde{\alpha}_{H,4,2}\mathrm{tr}(\tau^3X_R^{\mu})\mathrm{tr}(\tau^3X_R^{\nu})\mathrm{tr}(\tau^3X_{L,\nu})
+\tilde{\alpha}_{H,4,3}\mathrm{tr}(\tau^3X_L^{\mu})\mathrm{tr}(\tau^3X_R^{\nu})\mathrm{tr}(\tau^3X_{R,\nu})\nonumber\\
&&+\tilde{\alpha}_{H,4,4}\mathrm{tr}(\tau^3X_L^{\mu})\mathrm{tr}(\tau^3X_L^{\nu})\mathrm{tr}(\tau^3X_{R,\nu})
+i\tilde{\alpha}_{H,5,1}\mathrm{tr}(\tau^3X_{R,\nu})\mathrm{tr}(\tau^3\overline{W}_L^{\mu\nu})\nonumber\\
&&+i\tilde{\alpha}_{H,5,2}\mathrm{tr}(\tau^3X_{L,\nu})\mathrm{tr}(\tau^3\overline{W}_R^{\mu\nu})\}
+(\partial_{\mu}h)(\partial_{\nu}h)\tilde{\alpha}_{H,9}\mathrm{tr}(\tau^3X_R^\mu)\mathrm{tr}(\tau^3X_L^\nu)\nonumber\\
&&+(\partial_{\mu}h)^2\tilde{\alpha}_{H,11}\mathrm{tr}(\tau^3X_{L,\nu})\mathrm{tr}(\tau^3X_R^\nu)]\;.
\end{eqnarray}
Above interaction terms already include all possible $p^4$ order
CP-conserving and CP-violating operators and all $\alpha$
coefficients are functions of higgs field $h$. Left-right symmetry
will be explicitly realized for the theory if all coefficients with
subscript $_L$ are equal to their right hand partners denoted with
subscript $_R$. If they are not equal to each other, the left-right
symmetry is spontaneously violated by some underlying dynamics and
the differences between left and right coefficients then
characterize the strength of left-right symmetry violation .
Starting from above Lagrangian, we can read out various vertices
among electroweak gauge bosons which enable us to discuss
corresponding physical processes and further continue
 phenomenological researches. Theoretically,  we can
integrate out heavy right hand fields to see their effects to
ordinary EEWCL coefficients, or we can estimate the size of the
coefficients from existing models.

Now we first discuss two point vertices for gauge boson fields.
Taking unitary gauge $U_L=U_R=1$ and higgs field be its vacuum
expectation value $h=v$, the CP-conserving part of kinetic terms for
$W_L,W_R$ and $B$ become
\begin{eqnarray}
\mathcal{L}_{K,2}&=&-\frac{1}{4}(\partial_{\mu}W_{L,\nu}^a-\partial_{\nu}W_{L,\mu}^a)^2
-\frac{1}{4}(\partial_{\mu}W_{R,\nu}^a-\partial_{\nu}W_{R,\mu}^a)^2
-\frac{1}{4}(\partial_{\mu}B_{\nu}-\partial_{\nu}B_{\mu})^2\nonumber\\
&&+\frac{1}{2}g(\partial^{\mu}B^{\nu}-\partial^{\nu}B^{\mu})[
\alpha_{L,1}g_L(\partial_{\mu}W_{L,\nu}^3-\partial_{\nu}W_{L,\mu}^3)+\alpha_{R,1}g_R
(\partial_{\mu}W_{R,\nu}^3-\partial_{\nu}W_{R,\mu}^3)]\nonumber\\
&&+\frac{1}{4}\alpha_{L,8}g_L^2(\partial^{\mu}W_L^{3,\nu}-\partial^{\nu}W_L^{3,\mu})^2
+\frac{1}{4}\alpha_{R,8}g_R^2(\partial^{\mu}W_R^{3,\nu}-\partial^{\nu}W_R^{3,\mu})^2\nonumber\\
&&+\frac{1}{4}\tilde{\alpha}_8g_Lg_R(\partial^{\mu}W_L^{3,\nu}-\partial^{\nu}W_L^{3,\mu})
(\partial_{\mu}W_{R,\nu}^3-\partial_{\nu}W_{R,\mu}^3)\;.
\end{eqnarray}
With convention $W^{^1_2}_{i,\mu}=\frac{1}{\sqrt{2}}(W^+_{i,\mu}\pm
W^-_{i,\mu})$, take orthogonal rotation $V$ and some scale
transformation to diagonalize the mixing among
$W_{L,\mu}^3,W_{R,\mu}^3,B_{\mu}$ by
\begin{eqnarray}
\left(\begin{array}{c}W_{L,\mu}^3\\ W_{R,\mu}^3\\
B_\mu\end{array}\right)=V\left(\begin{array}{ccc}\frac{1}{\sqrt{\lambda_1}}&0&0\\
0&\frac{1}{\sqrt{\lambda_2}}&0\\0&0&\frac{1}{\sqrt{\lambda_3}}\end{array}\right)\left(\begin{array}{c}
W_{L,\mu}^{3\prime}\\
W_{R,\mu}^{3\prime}\\
B_{\mu}'\end{array}\right)~~~~~
\end{eqnarray}
with orthogonal matrix $V$ and three eigenvalues
$\lambda_1,\lambda_2,\lambda_3$ satisfy
\begin{eqnarray}
\left(\begin{array}{ccc}1-\alpha_{L,8}g_L^2&-\frac{1}{2}\tilde{\alpha}_8g_Lg_R&-\alpha_{L,1}g_Lg\\
-\frac{1}{2}\tilde{\alpha}_8g_Lg_R&1-\alpha_{R,8}g_R^2&-\alpha_{R,1}g_Rg\\-\alpha_{L,1}g_Lg&-\alpha_{R,1}g_Rg&1
\end{array}\right)
=V\left(\begin{array}{ccc}\lambda_1&0&0\\
0&\lambda_2&0\\0&0&\lambda_3\end{array}\right)V^T\;.
\end{eqnarray}
One can check that up to order of $p^4$, the third eigenvalue is
$\lambda_3=1$ and the first and second eigenvalues are
$\lambda_{\pm}$ or $\lambda_{\mp}$ which are correspond to two
phases of the theory
\begin{eqnarray}
\lambda_{\pm}=1\!-\frac{1}{2}\alpha_{L,8}g_L^2\!-\frac{1}{2}\alpha_{R,8}g_R^2
           \pm[\alpha_{L,1}^2g_L^2g^2\!+\!\alpha_{R,1}^2g_R^2g^2
           \!+\frac{1}{4}\tilde{\alpha}_8^2g_L^2g_R^2\!+\frac{1}{4}(\alpha_{L,8}g_L^2
           \!-\!\alpha_{R,8}g_R^2)^2]^{1/2}\;,~~~~
\end{eqnarray}
then kinetic term are normalized to standard form
\begin{eqnarray}
\mathcal{L}_{K,2}&=&-\frac{1}{2}(\partial_{\mu}W_{L,\nu}^+-\partial_{\nu}W_{L,\mu}^+)(\partial^{\mu}W_L^{-,\nu}-\partial^{\nu}W_L^{-,\mu})
-\frac{1}{2}(\partial_{\mu}W_{R,\nu}^+-\partial_{\nu}W_{R,\mu}^+)(\partial^{\mu}W_R^{-,\nu}-\partial^{\nu}W_R^{-,\mu})\nonumber\\
&&-\frac{1}{4}(\partial_{\mu}W_{L,\nu}^{3\prime}-\partial_{\nu}W_{L,\mu}^{3\prime})^2
-\frac{1}{4}(\partial_{\mu}W_{R,\nu}^{3\prime}-\partial_{\nu}W_{R,\mu}^{3\prime})^2
-\frac{1}{4}(\partial_{\mu}B_{\nu}'-\partial_{\nu}B_{\mu}')^2
\end{eqnarray}
The mass terms from our chiral Lagrangian can be read out as
 \begin{eqnarray} \mathcal{L}_M
&=&\frac{1}{2}f_L^2g_L^2W^+_{L,\mu}W^{-,\mu}_L+\frac{1}{2}f_R^2g_R^2W^+_{R,\mu}W^{-,\mu}_R
-\frac{1}{2}\kappa f_Lf_Rg_Lg_R(W^+_{L,\mu}W^{-,\mu}_R+W^+_{R,\mu}W^{-,\mu}_L)\nonumber\\
&&+\frac{1}{4}(1-\beta_{L,1})f_L^2(g_LW^3_{L,\mu}-gB_{\mu})^2
 +\frac{1}{4}(1-\beta_{R,1})f_R^2(g_RW^3_{R,\mu}-gB_{\mu})^2\nonumber\\
&&-\frac{1}{2}(\kappa+\tilde{\beta}_1)
f_Lf_R(g_LW^3_{L,\mu}-gB_{\mu})(g_RW^{3,\mu}_R-gB^{\mu})\;.
\end{eqnarray}
Take orthogonal rotation for $W^{\pm}_L$ and $W^{\pm}_R$ as
\begin{eqnarray}
\left(\begin{array}{c}W^{\pm}_L\\W^{\pm}_R\end{array}\right)=\left(\begin{array}{cc}\cos\xi&\sin\xi\\-\sin\xi&\cos\xi
\end{array}\right)=\left(\begin{array}{c}W^{\pm}_1\\W^{\pm}_2\end{array}\right)\hspace{2cm}
\tan 2\xi=\frac{2\kappa
f_Lf_Rg_Lg_R}{f_L^2g_L^2-f_R^2g_R^2}\label{xiresult}
\end{eqnarray}
and orthogonal rotation $\tilde{V}$ to diagonalize the mixing among
$W_{L,\mu}^3,W_{R,\mu}^3,B_{\mu}$ by
\begin{eqnarray}
&&\hspace{-0.5cm}\left(\begin{array}{c}W_{L,\mu}^3\\ W_{R,\mu}^3\\
B_\mu\end{array}\right)=V\Lambda\tilde{V}\left(\begin{array}{c}
Z_{1,\mu}\\
Z_{2,\mu}\\
A_{\mu}\end{array}\right)\hspace{0.5cm} \Lambda
V^T\tilde{M}_0^2V\Lambda=\tilde{V}\left(\begin{array}{ccc}
M_{Z_1}^2&0&0\\0&M_{Z_2}^2&0\\0&0&0\end{array}\right) \tilde{V}^T\hspace{0.5cm}
\Lambda\equiv\left(\begin{array}{ccc}\frac{1}{\sqrt{\lambda_1}}&0&0\\
0&\frac{1}{\sqrt{\lambda_2}}&0\\0&0&1\end{array}\right)\nonumber\\
&&\hspace{-1cm}\tilde{M}_0^2\!=\!\left(\begin{array}{ccc}\!\!\frac{1}{2}(1-\beta_{L,1})f_L^2g_L^2&
-\frac{1}{2}(\kappa\!+\!\tilde{\beta}_1) f_Lf_Rg_Lg_R&
\left[(\beta_{L,1}\!-\!1)f_L\!\!+\!(\kappa\!+\!\tilde{\beta}_1)f_R\right]\!\frac{f_Lg_Lg}{2}\!\!\!\\
\!\!-\frac{1}{2}(\kappa\!+\!\tilde{\beta}_1)
f_Lf_Rg_Lg_R&\frac{1}{2}(1-\beta_{R,1})f_R^2g_R^2&
\left[(\beta_{R,1}\!-\!1)f_R\!\!+\!(\kappa\!+\!\tilde{\beta}_1)f_L
\right]\!\frac{f_Rg_Rg}{2}\!\!\!\!\\
\!\!\left[(\beta_{L,1}\!\!-\!\!1)\!f_L\!\!+\!(\kappa\!+\!\!\tilde{\beta}_1)\!f_R\right]\!\frac{f_Lg_Lg}{2}
&\left[(\beta_{R,1}\!\!-\!\!1)\!f_R\!\!+\!(\kappa\!+\!\!\tilde{\beta}_1)\!f_L
\right]\!\frac{f_Rg_Rg}{2} &
\left[\frac{1\!-\!\beta_{L,1}}{2}\!f_L^2\!+\!\frac{1\!-\!\beta_{R,1}}{2}\!f_R^2\!-\!(\kappa\!+\!\!\tilde{\beta}_1)
f_Lf_R\right]\!g^2\!\!\!\end{array}\right)\;.\nonumber
\end{eqnarray}
The mass term then is diagonalized as
\begin{eqnarray} \mathcal{L}_M
&=&M^2_{W_1}W_1^{+\mu}W_{1,\mu}^-+M^2_{W_2}W_2^{+\mu}W_{2,\mu}^-+\frac{1}{2}M_{Z_1}^2Z_1^2+\frac{1}{2}M_{Z_2}^2Z_2^2
\end{eqnarray}
with masses
\begin{eqnarray}
M^2_{W_1}&=&\frac{1}{4}[f_L^2g_L^2+f_R^2g_R^2-\sqrt{(f_L^2g_L^2-f_R^2g_R^2)^2+4\kappa^2
f_L^2f_R^2g_L^2g_R^2}]
\approx\frac{1}{2}f_L^2g_L^2(1-\kappa^2)(1-\kappa^2\frac{f_L^2g_L^2}{f_R^2g_R^2})\;,\nonumber\\
M^2_{W_2}&=&\frac{1}{4}[f_L^2g_L^2+f_R^2g_R^2+\sqrt{(f_L^2g_L^2-f_R^2g_R^2)^2+4\kappa^2
f_L^2f_R^2g_L^2g_R^2}]
\approx\frac{1}{2}f_R^2g_R^2[1+\kappa^2\frac{f_L^2g_L^2}{f_R^2g_R^2}]\,\label{Mresult}\\
M^2_{Z_1}&=&
\frac{f_L^2[(1-\beta_{L,1})(1-\beta_{R,1})-(\kappa+\tilde{\beta}_1)^2]}{2(1-\beta_{R,1})(g_R^2+g^2)}
  \times  [g_R^2g^2+g_L^2g_R^2+g_L^2g^2\nonumber\\
&&-2\alpha_{L,1}(g_R^2+g^2)g_L^2g_R^2g^2+2g_R^4g^4\alpha_{R,1}
    +\alpha_{L,8}g_L^4(g_R^2+g^2)^2+\alpha_{R,8}g_R^4g^4-\alpha_8(g_R^2+g^2)g_L^2g_R^2g^2]\nonumber\\
     M^2_{Z_2}&=&
    \frac{1}{2}[(1-\beta_{R,1})f_R^2(g_R^2+g^2)-2f_Lf_Rg^2(\kappa+\beta_1)+(1-\beta_{L,1})f_L^2(g_L^2+g^2)]\nonumber\\
&&-\alpha_{L,1}g^2g_L^2[f_L^2(1-\beta_{L,1})-f_Lf_R(\kappa+\beta_1)]
  -\alpha_{R,1}g^2g_R^2[f_R^2(1-\beta_{R,1})-f_Lf_R(\kappa+\beta_1)]\nonumber\\
&&+\frac{1}{2}\alpha_{L,8}g_L^4f_L^2(1-\beta_{L,1})+\frac{1}{2}\alpha_{R,8}g_R^4f_R^2(1-\beta_{R,1})
  -\frac{1}{2}\alpha_8g_L^2g_R^2f_Lf_R(\kappa+\beta_1)-M^2_{Z_1}\nonumber\;.
\end{eqnarray}
where to reduce the length of the formulae, $M^2_{Z_1}$ and
$M^2_{Z_2}$ are only accurate up to order of $f_L^3/f_R$. Note all
$\alpha$, $\beta$, $\kappa$ and $f$ coefficients appeared in
(\ref{xiresult}) and (\ref{Mresult}) are their values in $h=v$.

 The right hand gauge bosons are expected
to be heavy and below their thresholds the interactions among
remaining left hand gauge bosons should be described by conventional
EWCL for which the effective interactions will receive the
contribution from right hand gauge bosons, we can compute these
contributions by integrating out right hand gauge boson fields. In
the following, we make the lowest order estimations by just
subtracting out right hand gauge boson fields with their classical
solution of tree order field equations. This treatment ignores the
loop contributions and only include in effects from exchanging the
right hand gauge boson at tree Feynman diagrams. Loop effects
usually are suppressed by a factor $1/16\pi^2$ and will be discussed
in future.

Before dealing with right hand gauge boson fields, we first handle
three right hand goldstone bosons by simply taking unitary gauge
$U_R=I$. Note this unitary gauge can always be realized by taking
suitable right hand gauge transformations. Then the $p^2$ order
equations of motion for $W_{R,\mu}^{1,2}$ give solution
$ig_RW^{1,2}_{R,\mu}=\delta_1X^{1,2}_{L,\mu}$ and equation for
$W_{R,\mu}^3$ give solution
$ig_RW^3_{R,\mu}=igB_{\mu}+\delta_2tr[\tau^3X_{L,\mu}]$ with
\begin{eqnarray}
\delta_1\equiv\kappa\frac{f_L}{f_R}\;,\hspace{2cm}
\delta_2\equiv\frac{(2\tilde{\beta}_1+\kappa)f_Lf_R}{(f_R^2-2\beta_{R,1}f_L^2)}\;.
\end{eqnarray}
Substitute the solution back to our chiral Lagrangian
$\mathcal{L}_2+\mathcal{L}_4$, after some algebra, we recover
standard EEWCL developed in Ref.\cite{EWCL} with expression
\begin{eqnarray}
\mathcal{L}_\mathrm{low~energy}&=&-V(h)
+\frac{1}{2}(\partial_{\mu}h)^2-\frac{1}{4}f^2{\rm
tr}(X_{L,\mu}^2)+\frac{1}{4}\beta_1f^2[{\rm
tr}(\tau^3X_{L,\mu})]^2-\frac{1}{4}W_{L,\mu\nu}^aW_L^{\mu\nu,a}\nonumber\\
&&-\frac{1}{4}K_BB_{\mu\nu}^2 +\frac{1}{2}\alpha_1gB_{\mu\nu}{\rm
tr}(\tau^3\overline{W}_i^{\mu\nu}) +i\alpha_2gB_{\mu\nu}{\rm
tr}(\tau^3X^{\mu}_LX^{\nu}_L)
 +2i\alpha_3{\rm tr}(\overline{W}_{L,\mu\nu}X^{\mu}_LX^{\nu}_L)
\nonumber\\
 && +\alpha_4[{\rm tr}(X_{L,\mu}X_{L,\nu})]^2+\alpha_5[{\rm tr}(X_{L,\mu}^2)]^2
 +\alpha_6{\rm tr}(X_{L,\mu}X_{L,\nu}){\rm tr}(\tau^3X^{\mu}_L){\rm tr}(\tau^3X^{\nu}_L)
 \nonumber\\
 &&+\alpha_7{\rm tr}(X_{L,\mu}^2)[{\rm tr}(\tau^3X_{L,\nu})]^2
 +\frac{1}{4}\alpha_8[{\rm tr}(\tau^3\overline{W}_{L,\mu\nu})]^2
 +i\alpha_9{\rm tr}(\tau^3\overline{W}_{L,\mu\nu}){\rm tr}(\tau^3X_L^{\mu}X_L^{\nu})
 \nonumber\\
 &&+\frac{1}{2}\alpha_{10}[{\rm tr}(\tau^3X_{L,\mu}){\rm tr}(\tau^3X_{L,\nu})]^2+
 \alpha_{11}\epsilon^{\mu\nu\rho\lambda}{\rm tr}(\tau^3X_{L,\mu}){\rm tr}(X_{L,\nu}\overline{W}_{L,\rho\lambda})
 \nonumber\\
&&+2\alpha_{12}{\rm tr}(\tau^3X_{L,\mu}){\rm
tr}(X_{L,\nu}\overline{W}_L^{\mu\nu})+\frac{1}{4}\alpha_{13}g\epsilon^{\mu\nu\rho\sigma}\!B_{\mu\nu}{\rm
tr}(\tau^3\overline{W}_{L,\rho\sigma}\!)
\nonumber\\
&&+\frac{1}{8}\alpha_{14}\epsilon^{\mu\nu\rho\sigma}{\rm
tr}(\tau^3\overline{W}_{L,\mu\nu}){\rm
tr}(\tau^3\overline{W}_{L,\rho\sigma})+(\partial_{\mu}h)\{\alpha_{H,1}\mathrm{tr}(\tau^3X_L^\mu)
\mathrm{tr}(X_{L,\nu}^2)\nonumber\\
&&+\alpha_{H,2}\mathrm{tr}(\tau^3X_L^\nu)\mathrm{tr}(X_L^{\mu}X_{L,\nu})+
\alpha_{H,3}\mathrm{tr}(\tau^3X_L^\nu)\mathrm{tr}(\tau^3X_L^{\mu}X_{L,\nu})\nonumber\\
&&+\alpha_{H,4}\mathrm{tr}(\tau^3X_L^{\mu})[\mathrm{tr}(\tau^3X_{L,\nu})]^2
+i\alpha_{H,5}\mathrm{tr}(\tau^3X_{L,\nu})\mathrm{tr}(\tau^3\overline{W}_L^{\mu\nu})
+ ig\alpha_{H,6}B^{\mu\nu}\mathrm{tr}(\tau^3X_{L,\nu})\nonumber\\
&& +i\alpha_{H,7}\mathrm{tr}(\tau^3\overline{W}_L^{\mu\nu}X_{L,\nu})
+i\alpha_{H,8}\mathrm{tr}(\overline{W}_L^{\mu\nu}X_{L,\nu})\}
+(\partial_{\mu}h)(\partial_{\nu}h)[\alpha_{H,9}\mathrm{tr}(\tau^3X_L^\mu)\mathrm{tr}(\tau^3X_L^\nu)\nonumber\\
&&+\alpha_{H,10}\mathrm{tr}(X_L^{\mu}X_L^\nu)\}
+(\partial_{\mu}h)^2\{\alpha_{H,11}[\mathrm{tr}(\tau^3X_{L,\nu})]^2+
\alpha_{H,12}\mathrm{tr}(X_{L,\nu}^2)\}\nonumber\\
&&+\alpha_{H,13}(\partial_{\mu}h)^2(\partial_{\nu}h)\mathrm{tr}(\tau^3X_L^\nu)
+\alpha_{H,14}(\partial_{\mu}h)^4
\end{eqnarray}
$p^2$ order coefficients expressed in terms of those parameters
appeared in $\mathcal{L}_2$ as follows
\begin{eqnarray}
f^2&=&f_L^2(1-3\kappa^2)\;,\\
\beta_1&=&\frac{\beta_{L,1}+\beta_{R,1}\tilde{\delta}_2^2+(2\tilde{\beta}_1+\kappa)\tilde{\delta}_2
-\tilde{\delta}_2^2-\frac{1}{2}\kappa^2}{1-3\kappa^2}\stackrel{f_L\ll
f_R}{--\rightarrow}
\frac{\beta_{L,1}+\beta_{R,1}(2\tilde{\beta}_1+\kappa)^2-\frac{1}{2}\kappa^2}{1-3\kappa^2}\;.~~~
\end{eqnarray}
with parameter
$\tilde{\delta}_2\equiv\frac{(2\tilde{\beta}_1+\kappa)f_R^2}{(f_R^2-2\beta_{R,1}f_L^2)}$.
 Further computations give all $p^4$ order coefficients of EWCL from $\mathcal{L}_2$ and
$\mathcal{L}_4$,
\begin{eqnarray}
K_B&=&1+(\frac{1}{2}\alpha_{R,1}+\frac{1}{4}\alpha_{R,8})(1+\delta_1-2\delta_2)\nonumber\\
\alpha_1&=&\alpha_{L,1}+\alpha_{R,1}\delta_2+\alpha_{R,8}\delta_2
        +\frac{1}{2}\tilde{\alpha}_8(1+\delta_1-2\delta_2)\;,\nonumber\\
\alpha_2&=&\alpha_{L,2}-\alpha_{R,1}(\delta_1-2\delta_2)
        -\alpha_{R,8}(\delta_1-2\delta_2)+\tilde{\alpha}_{2}\delta_1
        +(\tilde{\alpha}_{3,2}+\tilde{\alpha}_{9,2})(1+\delta_1-2\delta_2)\nonumber\\
        &&+\tilde{\alpha}_{3,4}\delta_1+\tilde{\alpha}_{9,4}\delta_1\;,\nonumber\\
\alpha_3&=&\alpha_{L,3}+\tilde{\alpha}_{3,2}\delta_1
        +\tilde{\alpha}_{3,3}\frac{\delta_2}{2}-\tilde{\alpha}_{3,4}(\frac{\delta_1}{2}-\frac{\delta_2}{2})\;,\nonumber\\
\alpha_4&=&\alpha_{L,4}+\tilde{\alpha}_{4,5}\delta_1+4\tilde{\alpha}_{3,2}(\delta_1-\delta_2)
        +\tilde{\alpha}_{9,2}(2\delta_1-4\delta_2)-4\tilde{\alpha}_{3,2}\delta_1\;,\nonumber\\
\alpha_5&=&\alpha_{L,5}+\tilde{\alpha}_{5,4}\delta_1-4\tilde{\alpha}_{3,2}(\delta_1-\delta_2)
        -\tilde{\alpha}_{9,2}(2\delta_1-4\delta_2)+4\tilde{\alpha}_{3,2}\delta_1\;,\nonumber\\
\alpha_6&=&\alpha_{L,6}+\tilde{\alpha}_{4,5}(\frac{\delta_2}{2}-\frac{\delta_1}{2})+\tilde{\alpha}_{6,6}\delta_1
        +\tilde{\alpha}_{6,7}\delta_2-4\tilde{\alpha}_{3,2}(\delta_1-\delta_2)
        -\tilde{\alpha}_{9,2}(2\delta_1-4\delta_2)\;,\nonumber\\
\alpha_7&=&\alpha_{L,7}+\tilde{\alpha}_{5,4}(\frac{\delta_2}{2}-\frac{\delta_1}{2})+\tilde{\alpha}_{7,5}\delta_1
        +\tilde{\alpha}_{7,6}\delta_2+4\tilde{\alpha}_{3,2}(\delta_1-\delta_2)
        +\tilde{\alpha}_{9,2}(2\delta_1-4\delta_2)\;,\nonumber\\
\alpha_8&=&\alpha_{L,8}+\tilde{\alpha}_8\delta_2\;,\nonumber\\
\alpha_9&=&\alpha_{L,9}+\tilde{\alpha}_{9,3}\delta_1+\tilde{\alpha}_{9,2}\delta_2
        -\frac{1}{4}\tilde{\alpha}_8(2\delta_1-4\delta_2)+\tilde{\alpha}_{3,2}(-\delta_1+\delta_2)
        -\tilde{\alpha}_{3,3}\frac{\delta_2}{2}+\tilde{\alpha}_{3,4}(\frac{\delta_1}{2}-\frac{\delta_2}{2})\;,\nonumber\\
\alpha_{10}&=&\alpha_{L,10}+\tilde{\alpha}_{6,6}(\delta_2-\delta_1)+\tilde{\alpha}_{7,5}(\delta_2-\delta_1)\;,\nonumber\\
\alpha_{11}&=&\alpha_{L,11}
        +\tilde{\alpha}_{11,2}\delta_2+\tilde{\alpha}_{11,5}\delta_1
        +\tilde{\alpha}_{11,3}\delta_1\;,\nonumber\\
\alpha_{12}&=&\alpha_{L,12}
        +\tilde{\alpha}_{12,2}\delta_2+\tilde{\alpha}_{12,5}\delta_1\;,\nonumber\\
\alpha_{13}&=&\alpha_{L,13}+\alpha_{R,13}(\delta_1-\delta_2)+\alpha_{R,14}(\delta_1-\delta_2)
        +4\alpha_{R,15}\delta_2-\frac{1}{8}\tilde{\alpha}_{14}(2\delta_1-4\delta_2)\nonumber\\
        &&+2\tilde{\alpha}_{15}\delta_2+(\frac{1}{2}\tilde{\alpha}_{14}+2\tilde{\alpha}_{15})(1+\delta_1-2\delta_2)\;,\nonumber\\
\alpha_{14}&=&\alpha_{L,14}+\tilde{\alpha}_{14}\delta_2+4\tilde{\alpha}_{15}(-\delta_1+\delta_2)\;,\\
\alpha_{H,1}&=&\delta_2\tilde{\alpha}_{H,1,1}\;,\nonumber\\
\alpha_{H,2}&=&\delta_2\tilde{\alpha}_{H,2,1}\;,\nonumber\\
\alpha_{H,3}&=&\delta_2\tilde{\alpha}_{H,3,1}+\tilde{\alpha}_{H,5,2}(2\delta_1-4\delta_2)\;,\nonumber\\
\alpha_{H,5}&=&\tilde{\alpha}_{H,5,2}\delta_2\;,\nonumber\\
\alpha_{H,6}&=&\delta_2\bar{\alpha}_{HR,5}+g\tilde{\alpha}_{H,5,2}(1+\delta_1-2\delta_2)
        +g\delta_2\bar{\alpha}_{HR,6}+\delta_2\tilde{\alpha}_{H,5,1}
        +\bar{\alpha}_{HR,8}\delta_1\frac{g}{2}\;,\nonumber\\
\alpha_{H,9}&=&\delta_2\tilde{\alpha}_{H,9,1}\;,\nonumber\\
\alpha_{H,4}&=&\alpha_{H,7}=\alpha_{H,8}=\alpha_{H,10}=\alpha_{H,11}=\alpha_{H,12}=0\;,\nonumber\\
\alpha_{H,13}&=&\delta_2\bar{\alpha}_{HR,13}\nonumber
\end{eqnarray}
where to avoid the lengthy expressions, we ignore terms of order
$\alpha\delta^2$ and $\alpha\delta^3$ in $p^4$ order results. In
terms of above results, we can read out corrections to well known
parameter S, T, U \cite{STU} from general right hand gauge boson
fields with help of their relation to $\alpha_1,\beta_1,\alpha_8$
given in Ref.\cite{EWCL},
\begin{eqnarray}
&&\Delta
S=-16\pi\Delta\alpha_1=-16\pi[\alpha_{R,1}\delta_2+\alpha_{R,8}\delta_2
        +\frac{1}{2}\tilde{\alpha}_8(1+\delta_1-2\delta_2)]\;,\\
&&\Delta \alpha_{\rm em} T=2\Delta\beta_1=
\frac{2\beta_{R,1}\tilde{\delta}_2^2+2(2\tilde{\beta}_1+\kappa)\tilde{\delta}_2
-2\tilde{\delta}_2^2-\kappa^2}{1-3\kappa^2}\;,\\
&&\Delta
U=-16\pi\Delta\alpha_8=-16\pi\tilde{\alpha}_8\delta_2\;.~~~~~
\end{eqnarray}
 We see, if right hand gauge boson is much heavier than the left
hand ones $f_R\gg f_L$, $S$ and $T$ parameters will receive
corrections of $-8\pi\tilde{\alpha}_8$ and
$[2\beta_{R,1}(2\tilde{\beta}_1+\kappa)^2-\kappa^2]/\alpha_{\rm
em}(1-3\kappa^2)$, while $U$ parameter will have no correction. In
this approximation of $f_R\gg f_L$, for general triple vertices
among electroweak gauge fields
\begin{eqnarray}
\frac{{\cal
L}_{WWV}}{g_{WWV}}&=&ig_1^V(W_{\mu\nu}^+W^{-\mu}V^{\nu}-W_{\mu\nu}^-W^{+\mu}V^{\nu})+i\kappa_VW^+_{\mu}W^-_{\nu}V^{\mu\nu}
-g_4^VW^+_{\mu}W^-_{\nu}(\partial^{\mu}V^{\nu}+\partial^{\nu}V^{\mu})\nonumber\\
&&+g_5^V\epsilon^{\mu\nu\rho\lambda}[W^+_{\mu}(\partial_{\rho}W^-_{\nu})-(\partial_{\rho}W^+_{\mu})W^-_{\nu}]V_{\lambda}
+i\tilde{\kappa}_VW^+_{\mu}W^-_{\nu}\tilde{V}^{\mu\nu}\;,
\end{eqnarray}
 with the relation between anomalous couplings of
triple vertices and EWCL coefficients given in Ref.\cite{EWCL}, we
find following constraints for corrections from right hand gauge
boson fields,
\begin{eqnarray}
\Delta g_1^Z\!=\Delta\kappa_Z\!+\!\frac{s^2}{c^2}\Delta\kappa_\gamma
=\frac{\frac{1}{2}\alpha_{\rm em}\Delta T-\frac{e^2}{16\pi
c^2}\Delta S}{c^2-s^2} \hspace{0.7cm} \Delta g_5^Z=\Delta
g_4^Z=0\hspace{0.7cm} c^2\Delta\tilde{\kappa}_Z
=-s^2\Delta\tilde{\kappa}_{\gamma}~~~~~
\end{eqnarray}
Further, with same approximation, the right hand gauge boson fields
contribute to general anomalous quartic vertices with
\begin{eqnarray}&&\hspace{-0.5cm}\Delta{\cal
L}^{\rm QGV}=\Delta g_1^Z\{\frac{e^{*2}c^2}{4\sin^2\theta_W}
[W^+_{\mu}W^{+\mu}W^-_{\nu}W^{-\nu}-(W^+_{\mu}W^{-\mu})^2]
+2e^{*2}\cot^2\theta_W[W^+_{\mu}Z^{\mu}W^-_{\nu}Z^{\nu}\nonumber\\
&&\hspace{1.4cm}-W^+_{\mu}W^{-\mu}Z_{\nu}Z^{\nu}]+e^{*2}\cot\theta_W
[(W^+_{\mu}W^-_{\nu}+W^+_{\nu}W^-_{\mu})Z^{\mu}A^{\nu}-W^+_{\mu}W^{-\mu}Z_{\nu}A^{\nu}]\}\;.~~~~
\end{eqnarray}

In general, we find the contribution of our chiral Lagrangian for
left-right symmetric models to EWCL coefficient $\alpha_i$ can be
divided into three parts: the first is the primary term
$\alpha_{L,i}$ which comes from original left hand gauge boson
interaction; the second is the term independent of $\delta$ and
linear in $\alpha$, only
 $\alpha_1,\alpha_2,\alpha_{13}$ receive such kind corrections with
 values of
 $\frac{1}{2}\tilde{\alpha}_8,\tilde{\alpha}_{3,2}+\tilde{\alpha}_{9,2},
\frac{1}{2}\tilde{\alpha}_{14}+2\tilde{\alpha}_{15}$ respectively
and all other coefficients do not have such kind terms; the third
term is the term not only linear in $\alpha$, but also proportional
to $\delta$. Since $\delta$ is order of $f_L/f_R$ which is a small
 number when right hand interaction scale is heavier than the left hand interaction scale,
 the third term is smaller in orders than the second term.
Although due to lack detail information on size of all those
parameters in $\mathcal{L}_2$ and $\mathcal{L}_4$, we can not
estimate their contributions to EWCL coefficients quantitatively,
qualitatively we can judge that only $\alpha_1,\alpha_2,\alpha_{13}$
receive relatively large contributions of order $\alpha$ from
exchanging virtual right hand gauge bosons, all other coefficients
will only receive much smaller corrections of order
$\alpha*f_L/f_R$.

 To summarize, we have set
up the most general electroweak chiral Lagrangian for left-right
symmetric models up to order of $p^4$ and discuss the gauge boson
masses and mixings. The contributions to conventional EWCL
coefficients from right hand gauge boson as virtual particle are
estimated.

\section*{Acknowledgments}

This work was  supported by National  Science Foundation of China
(NSFC) under Grant No. 10435040 and Specialized Research Fund for
the Doctoral Program of High Education of China.




\begin{thebibliography}{1}\label{biblio}

\bibitem{WangLiMing}
L-M.Wang, Q.Wang,  hep-ph/0605104

\bibitem{EWCL0}
T.Appelquist and C.Bernard, Phys. Rev. {\bf D22}, 200(1980);\\
A.Longhitano, Phys. Rev. {\bf D22}, 1166(1980); Nucl. Phys. {\bf
B188},118(1981)

\bibitem{EWCL}
 T.Appelquist and G-H. Wu, Phys. Rev. {\bf D48},
3235(1993); {\bf D51}, 240(1995)



\bibitem{LRSMs}
P.Duka, J.Gluza and M. Zralek,  Ann. Phys. {\bf 280}, 336(2000) and
references therein;\\
F.Siringa and Luca Marotta, Phys. Rev. {\bf D74}, 115001(2006) and
references therein.

\bibitem{He}
R.S.Chivukula, D.A.Dicus, H-J.He,  Phys. Lett. {\bf B525}, 175(2002)

\bibitem{STU}
M.E.Peskin and T.Takeuchi, Phys. Rev. Lett. {\bf 65}, 964(1990);
Phys. Rev. {\bf D46}, 381(1992)

\bibitem{2doublets}
B.Brahmachari, E.Ma and U.Sarkar, Phys. Rev. Lett. {\bf 91},
011801(2003);\\
F.M.L. Almeida, Jr., Y.A. Coutinho, J.A. Martins
Simoes, J. Ponciano, A.J. Ramalho, Stenio Wulck, M.A.B. Vale, Eur.
Phys. J. {\bf C38}, 115(2004)

\bibitem{EWCLmatter}
E.Bagan, D.Espriu and J.Manzano, Phys. Rev. {\bf D60}, 114035(1999)
\end{thebibliography}
\end{document}